\documentstyle[preprint,aps]{revtex}
\begin{document}

\draft         
\preprint{IFT-P.018/95 --- IFUSP-P 1145}
\title{ Excited Leptonic States in Polarized \boldmath{$e^-\gamma$}
and \boldmath{$e^+e^-$} Collisions }
\author{O.\ J.\ P.\ \'Eboli }
\address{Instituto de F\'{\i}sica,
Universidade de S\~ao Paulo, \\
C.P.\ 66318, 05389-970 S\~ao Paulo, Brazil}
\author{E.\ M.\ Gregores, J.\ C.\ Montero, S.\ F.\ Novaes,
and D.\ Spehler}
\address{Instituto de F\'{\i}sica Te\'orica, Universidade
Estadual Paulista \\
Rua Pamplona 145, 01405-900 S\~ao Paulo, Brazil}
%
\maketitle
\begin{abstract}
We analyze the capability of the next generation of linear
electron--positron colliders to unravel the spin and couplings of
excited leptons predicted by composite models.  Assuming that these
machines will be able to operate both in the $e^+e^-$ and $e^-\gamma$
modes, we study the effects of the excited electrons of spin
$\frac{1}{2}$ and $\frac{3}{2}$ in the reactions   $e^-\gamma
\rightarrow e^-\gamma$ and $e^+e^- \rightarrow \gamma\gamma$. We
show how the use of polarized beams is able not only to increase the
reach of these machines, but also to determine the spin and couplings of
the excited states.
\end{abstract}

\section{Introduction}

The standard model (SM) of the electroweak interactions explains
extremely well all the available experimental data \cite{schaile}.
Notwithstanding, it has some unpleasant features, such as the large
number of free parameters, the proliferation of fermionic generations,
and their complex pattern of masses and mixing angles. A rather
natural explanation for the existence of the fermionic generations is
that the known leptons and quarks are composite \cite{comp}, sharing
some common constituents (preons). In this sense, it is conceivable
that the SM is just the low energy limit of a more fundamental theory,
which is characterized by a large mass scale $\Lambda$.  In general,
composite models exhibit a rich spectrum which includes many new
states like the excitations of the known particles.

Since the existence of excited fermions is an undeniable signal for
new physics beyond the SM, there have been several direct searches for
these particles at different accelerators. At the CERN Large
Electron--Positron Collider (LEP), the experiments excluded the
existence of excited spin--$\frac{1}{2}$ fermions with mass up to $46$
GeV from the pair production search, and up to $90$ GeV from direct
single production for a scale of compositeness $\Lambda < 2.5$ TeV
\cite{rev:lep}. Moreover, a limit on the mass of an excited
electron of $M_{e^\ast} > 127$ GeV at $95\%$ of confidence level
was set from the measurement of the $e^+e^- \rightarrow
\gamma\gamma$ cross section \cite{rev:lep}. On the other hand, the
experiments at the DESY $ep$ collider HERA searched, in a model
independent way, for resonances in the $e\gamma$, $\nu W$, and $eZ$
systems \cite{hera}, however the LEP bounds on excited leptons
couplings are about one order of magnitude more stringent in the mass
region just below the $Z$ mass.

Up to now all the direct searches for compositeness have failed, and
we expect that the next generation of accelerators, working at higher
center--of--mass energies, will be able to further extend the search
for composite states. On the theoretical side, there have been
extensive studies on the possibility of unraveling the existence of
excited fermions in $pp$ \cite{pp,kuhn}, $e^+e^-$
\cite{kuhn,ele:pos,hag,lep:hera,e:gam}, and $ep$ \cite{hag,lep:hera}
collisions at higher energies.

A particularly interesting machine for analyzing the substructure of
the electron and its neutrino is the Next Linear $e^+e^-$ Collider
(NLC) that is been planned to operate with a center--of--mass energy
of at least $500$ GeV and an integrated luminosity around 10 fb$^{-1}$
\cite{pal}.  At the NLC, it will be possible to convert an electron beam
into a photon one focusing a laser on the electron beam. By Compton
scattering, high energy photons are produced along the electron
direction carrying away a large amount of the beam energy
\cite{las0,laser:pol}.  The laser backscattering mechanism will allow
the NLC to operate in three different modes, $e^+e^-$, $e^-\gamma$, and
$\gamma \gamma$, opening up an opportunity for a deeper search for
compositeness \cite{lep:hera,e:gam}. A nice feature of the $e^-\gamma$
mode of NLC is the possibility of searching for new excited charged
leptons as resonances in the $e^-\gamma$ scattering \cite{e:gam}.

In this work we analyze the deviations from the SM predictions of the
reactions $e^-\gamma \rightarrow e^-\gamma $ and $e^+e^-
\rightarrow \gamma \gamma$ due to the exchange of excited
spin--$\frac{1}{2}$ and spin--$\frac{3}{2}$ fermions.  In particular,
it is important to determine the spin of the fermionic excitations
since the allowed values of the spin can give hints about the
underlying preonic structure \cite{kuhn}. We perform a detailed study
of the experimental signatures of excited fermions exploring the
possibility of polarizing both the electron and laser beams, and we
point out the best strategy to unravel the spin and the chiral
couplings of the excited leptons to the usual particles.  We also
study the discovery limits on the new physics parameters and we show
that an $e^-\gamma$ collider can probe very large values of the
compositeness scale ($\Lambda$), {\em e.g.\/} for excited electron
masses of the order of 400 GeV we can explore up to $\Lambda \sim 200$
TeV.

The outline of this paper is the following. In Sec.\ \ref{lag}, we
introduce the effective Lagrangians describing the excited fermion
couplings and discuss the existing low energy constraints. The analysis
of the reaction $e^-\gamma \rightarrow e^-\gamma$ is contained in Sec.\
\ref{eg:proc}, where we also present the main ingredients of the laser
backscattering mechanism. Section \ref{ee:proc} exhibits the study of
the reaction $e^+e^- \rightarrow \gamma \gamma$ for LEP II and NLC
energies and our results are summarized in Sec.\ \ref{concl}. This paper
is supplemented with an appendix that contains the complete helicity
amplitudes for the processes under study, as well as for the decays of
the excited leptons.

\section{ Effective Lagrangians}
\label{lag}

Composite models cannot be analyzed perturbatively since the preons
interact strongly at the energy scales of interest. Instead, we must
rely on effective Lagrangians to describe the couplings of excited
states with ordinary fermions and vector bosons. We demanded that the
effective Lagrangians for the excited fermions are $CP$ conserving and
that they respect the $U(1)_{\text{em}}$ gauge invariance. We
considered a magnetic moment type coupling among the excited
spin--$\frac{1}{2}$ fermion ($\Psi^{\ast}_{1/2}$), the ground state
fermion ($\psi$), and the photon that is described by the effective
Lagrangian \cite{ren,cab,low}
\begin{equation}
{\cal L}^{1/2}_{\text{eff}} = \frac{e}{2\Lambda}
\bar{\Psi}^{\ast}_{1/2} \sigma^{\mu\nu} (A + B \gamma_5) \psi \;
F_{\mu\nu} + \text{h.c.} \; ,
\label{l:12}
\end{equation}
where $\Lambda$ is the compositeness scale and $F_{\mu\nu}$ is the
electromagnetic field strength tensor. For the coupling of
spin--$\frac{3}{2}$ excited states ($\Psi^{\ast \mu}_{3/2}$) to usual
fermions and photons, we adopted the lowest order $U(1)_{\text{em}}$
gauge-invariant effective Lagrangian that can be constructed out of
these fields \cite{kuhn,salin,3:2:coll}
\begin{equation}
{\cal L}^{3/2}_{\text{eff}} = \frac{e}{\Lambda}
\bar{\Psi}^{\ast \mu}_{3/2} \gamma^{\nu}
(C + D \gamma_5) \psi \; F_{\mu\nu} + \text{h.c.} \; ,
\label{l:32}
\end{equation}
where $\sigma_{\mu\nu} = {i\over 2}[\gamma_\mu ,\gamma_\nu]$. The
constants $A$, $B$, $C$, and $D$ are assumed to be real in order to
preserve CP invariance. In general, the effective Lagrangians
(\ref{l:12}) and (\ref{l:32}) can be embodied in a wider class of
models \cite{hag} that respects the $SU(2)_L \times U(1)_Y$ gauge
invariance, provided that the couplings are chosen conveniently.
Since we are interested only in electromagnetic transitions, we
ignored the possible coupling of the excited states with the weak
gauge bosons in the present work.

The above couplings of the excited leptons allow them to decay
predominantly into the ground state lepton through the emission
of a photon with  widths
\begin{eqnarray}
\Gamma_{1/2} &\simeq & \Gamma_{1/2} (L_{1/2} \to e \gamma) =
\frac{\alpha M_{1/2}^3}{\Lambda^2}\; ,
\nonumber \\
\Gamma_{3/2} & \simeq & \Gamma_{3/2} (L_{3/2} \to e \gamma) =
\frac{\alpha M_{3/2}^3}{4 \Lambda^2} \; .
\end{eqnarray}
In this work, we assumed that the branching ratio of these particles
into electron--photon pairs is equal to one.  The helicity amplitudes
for these decays are presented in the Appendix.

The above Lagrangians are constrained by the direct searches in
collider experiments, as well as by their effect in the low--energy
phenomenology. Excited fermions can contribute to atomic parity
violation, electron--deuteron scattering \cite{rgEllis}, and the
anomalous magnetic moment of leptons ($e$ and $\mu$) \cite{ren,g-2}.
For an arbitrary choice of the couplings $A$, $B$, $C$, and $D$ the
most stringent bounds on excited fermions are due to their contribution
to the $(g-2)$ of the muon. However, these limits can be softened if
we consider only chiral couplings, {\it i.e.} the new interactions
have either a right--handed (RH) or left--handed (LH) structure
\cite{ren}. In this case the contributions of the spin--$\frac{1}{2}$
and spin--$\frac{3}{2}$ excited states to $(g-2)$ is proportional to
$m_\mu^2/\Lambda^2 $, and the bound reads $\Lambda \gtrsim 800$ GeV.
In our numerical results, we have taken into account the
experimental LEP bound on the mass and coupling of the excited
fermions and  also the bound coming from the very precise
measurement of anomalous magnetic moments.

\section{Electron--Photon Collisions}
\label{eg:proc}

\subsection{Polarized Laser Backscattering Distribution Functions
and Cross Sections}

In a linear collider it is possible to transform an electron (positron)
beam into a intense $\gamma$ one through the process of laser
backscattering \cite{las0}.  This mechanism relies on the fact that
Compton scattering of energetic electrons by soft laser photons gives
rise to high energy photons, that are collimated in the direction of the
incident electron.  Another very powerful feature of the Compton
backscattering mechanism is the possibility of obtaining a high degree
of polarization for the backscattered photons by polarizing the incoming
electron and (or) laser beams. The backscattered photon distribution
function for polarized electron and laser beams is \cite{laser:pol}
\begin{equation}
F(x,\zeta; P_e, P_l)=\frac{2\pi \alpha ^2}{\zeta\,m^2\,\sigma_c}
\left[\frac{1}{1-x}+1-x-4r( 1- r ) - P_e P_l \; r \; \zeta \;
( 2r-1)(2-x ) \right] \; ,
\label{f:pol}
\end{equation}
where $P_e$ is the mean parent electron longitudinal polarization,
$P_l$ represents the laser photon circular polarization, and
$\sigma_c$ is the Compton cross section
\begin{equation}
\sigma _c = \sigma_c^0+ P_e P_l \; \sigma_c^1  \; ,
\label{sig:com}
\end{equation}
with
\begin{eqnarray}
\sigma_c^0 &=& \frac{2\pi \alpha ^2}{\zeta\,m^2}
\left[ \left( 1- \frac{4}{\zeta} - \frac{8}{\zeta^2} \right)
\ln \left( \zeta+1 \right) + \frac{1}{2} + \frac{8}{\zeta} -
\frac{1}{2\left( \zeta+1\right) ^2} \right] \; ,
\nonumber \\
\sigma_c^1 &=& \frac{2\pi \alpha ^2}{\zeta\,m^2}
\left[\left( 1+\frac 2\zeta\right) \ln \left( \zeta+1\right) -
\frac{5}{2} + \frac{1}{\zeta+1} -
\frac{1}{2\left( \zeta+1\right) ^2}
\right] \; .
\label{sig:01}
\end{eqnarray}
We  defined the variables
\begin{equation}
x = \frac{\omega}{ E} \; ,
\;\;\;\;\;\;\;\;
\zeta =\frac{4 E \omega_0}{m^2}  \; ,
\;\;\;\;\;\;\;\;
r = \frac{x}{\zeta (1-x)} \; ,
\end{equation}
where $m$ and $E$ are the electron mass and energy, $\omega_0$ is the
laser energy, and $\omega$ is the backscattered photon energy.  The
variable $x \leq x_{\text{max}} \equiv \zeta/(\zeta +1)$ represents
the fraction of the electron energy carried by the backscattered
photon and $r \leq 1$.  In our calculations, we assumed $\zeta=
2(1+\sqrt{2}) \simeq 4.83$ in order to maximize the backscattered
photon energy without spoiling the luminosity through $e^+e^-$ pair
creation by the interaction between laser and backscattered photons.

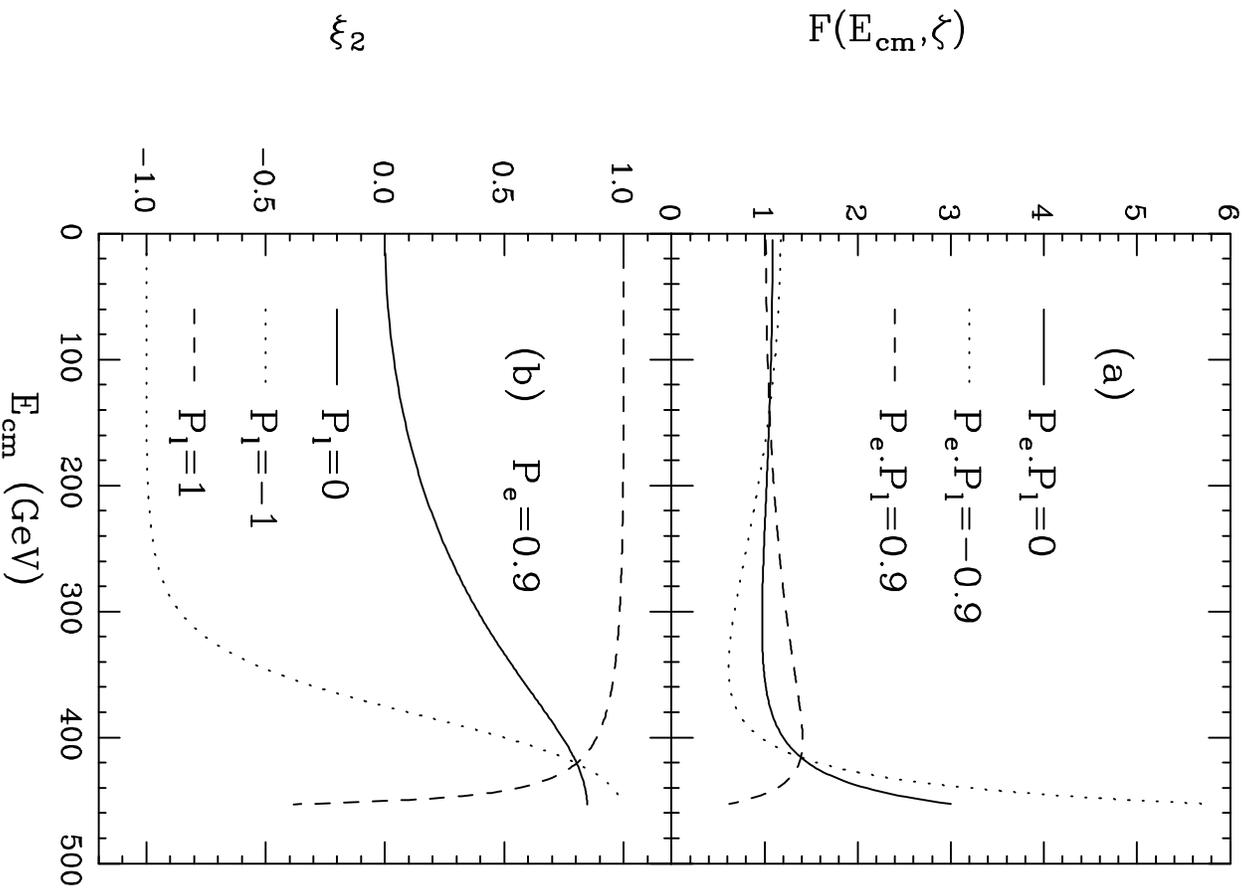
\begin{figure}
\protect
\caption{(a) Luminosity of backscattered photons in a $e^- \gamma$
collider as a function of the center--of--mass energy ($E_{\protect
\text{cm}}$). We assumed  $\protect\sqrt{s} = 500$ GeV, $\zeta = 4.83$,
and  $P_e \cdot P_l = 0$ (solid), $P_{e} \cdot P_l = -0.9$ (dots), and
$P_{e}\cdot P_l = 0.9$ (dashes). (b) Plot of $\xi_2$ as function of
$E_{\protect\text{cm}}$ for $\protect\sqrt{s} = 500$ GeV, $P_{e} = 0.9$,
and $P_l = 0$ (solid), $-1$ (dots), and $1$ (dashes).}
\label{polxi2}
\label{lback}
\end{figure}

The backscattered photon spectrum (\ref{f:pol}) depends only upon the
product $P_e P_l$, and the unpolarized distribution is recovered if
either the electron beam or the laser is not polarized. As can be
seen from Fig.\ \ref{lback}a, for negative values of this product the
spectrum is dominated by hard photons, otherwise it is quite broad. We
should notice that for $r=1/2$ the distribution function (\ref{f:pol})
depends on the electron and laser polarization only through $\sigma_c$
and the distribution functions, for any electron and laser
polarization, have approximately the same value at $x = \zeta / (\zeta
+2) \simeq 0.71$.

The mean backscattered photon helicity  is given by the Stokes
parameter
\begin{equation}
\xi_2 = \frac{P_e \,r\,\zeta\left[ 1+\left( 1-x\right)
\left( 2r-1\right) ^2\right] -P_l\left( 2r-1\right)
\left[ 1/(1-x) + 1 - x \right]}
{1/(1-x)+1-x-4r\left( 1-r\right)
- P_e P_l\,r\,\zeta\left( 2r-1\right) \left( 2-x\right) } \; .
\label{xi:2}
\end{equation}
For $x = x_{\text{max}}$ (or $r=1$) and $P_e =0$ or $P_l = \pm 1$, we
have $\xi_2 = - P_l$, {\it i.e.\/} the polarization of the
backscattered photon beam has the opposite value of the laser
polarization. Moreover, for $x = \zeta / (\zeta +2)$ (or $r=1/2$), the
Stokes parameter $\xi_2$ is independent of the laser polarization (see
Fig.\ \ref{lback}b) and is given by
\begin{equation}
\xi_2^{(r=1/2)} = P_e \frac{\zeta (\zeta +2)}{\zeta (\zeta +2) +4} \; .
\label{xi:2:r}
\end{equation}

The cross section for the reaction $e^- \gamma \rightarrow X$ in a $e^+
e^-$ linear collider, where the positron beam with longitudinal
polarization $P_p$ is converted into a backscattered photon beam, is
\begin{equation}
d\sigma_{P_e \xi_2} \left(e^- \gamma \rightarrow X \right) = \kappa
\int_{ x_{\text{min}}}^{x_{\text{max}}} dx \; F(x,\zeta; P_p, P_l) \;
d\hat{\sigma}_{P_e \xi_2} (e \gamma \rightarrow X) \; ,
\label{cross:pol}
\end{equation}
where $\kappa$ is the efficiency of the laser backscattering mechanism
\cite{foot1}, that we assumed to be one, and $d\hat{\sigma}_{P_e
\xi_2}$ is the polarized cross section for the subprocess $e^- \gamma
\rightarrow X$ which depends on $\hat{s} = x s$. In general, this
polarized cross section can be written as
\begin{eqnarray}
d\hat{\sigma}_{P_e \xi_2} &=&  \frac{1}{4}  \Bigl[(1+P_e \xi_2 )
\left(d\hat{\sigma}_ {++} + d\hat{\sigma}_ {--} \right) +  (P_e +
\xi_2 ) \left(d\hat{\sigma}_ {++} - d\hat{\sigma}_ {--} \right)
\nonumber \\
&+&   (1-P_e \xi_2 )\left(d\hat{\sigma}_ {+-} + d\hat{\sigma}_
{-+}\right)  +  (P_e - \xi_2  )\left(d\hat{\sigma}_ {+-} -
d\hat{\sigma}_{-+}\right)  \Bigr] \; ,
\label{elem:pol}
\end{eqnarray}
with $d\hat{\sigma}_{\lambda_e \lambda_{\gamma}}$
($\lambda_{e(\gamma)} = \pm 1$) being the polarized subprocess
cross section for full electron and photon polarizations, $P_e$ is
the  longitudinal polarization of the electron beam, and $\xi_2$
is given by Eq.\ (\ref{xi:2}).

\subsection{Results}

The complete set of the helicity amplitudes for the subprocess
\begin{equation}
e^- \gamma \rightarrow \left (e^-  , \;  L^-_{1/2(3/2)}\right )
\rightarrow e^-  \gamma \; ,
\label{proc}
\end{equation}
are presented in the Appendix for the exchange of spin--$\frac{1}{2}$
(\ref{amp:12:full}) and spin--$\frac{3}{2}$ (\ref{amp:32:full})
excited fermions in the $s$-channel.  In order to avoid the strong
bounds coming from the muon $(g-2)$ measurements, we assumed either
left--handed (LH) ($A=-B=1$ and $C=-D=1$) or right--handed (RH)
($A=B=1$ and $C=D=1$) couplings.
In order to quantify the potential of an $e^-\gamma$ collider to search
for excited states, we defined the statistical significance of the
signal (${\cal S}$) by
\begin{equation}
{\cal S} \equiv \frac{| \sigma_{\text{exc}} - \sigma_{\text{QED}}|}
{\sqrt{\sigma_{\text{QED}}}} \sqrt{\cal L} \; ,
\end{equation}
where $\sigma_{\text{exc}}$ ($\sigma_{\text{QED}}$) is the cross
section associated to the excited lepton (QED) contributions and
${\cal L}$ is the integrated luminosity of the machine.

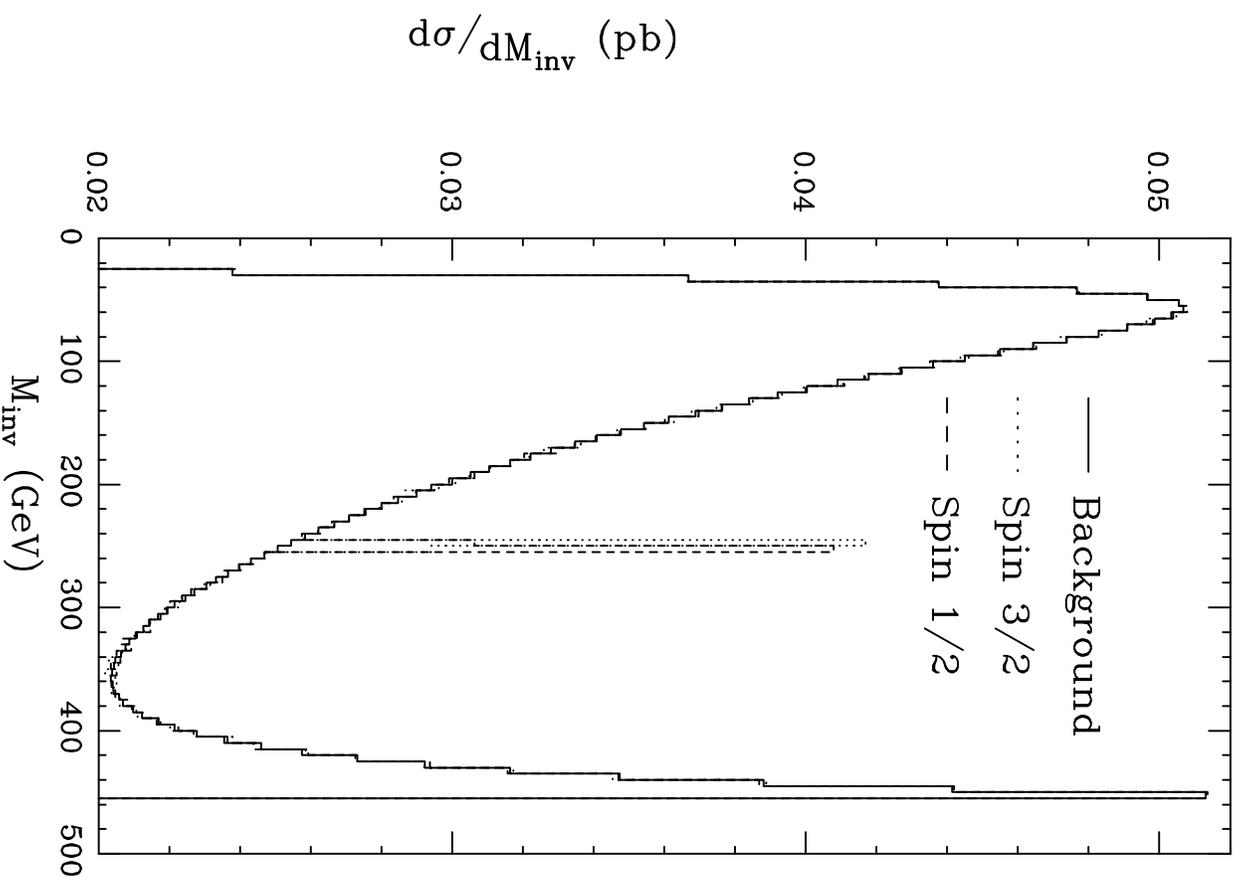
\begin{figure}
\protect
\caption{Invariant mass distribution for the process
(\protect\ref{proc}), for the exchange of spin--$\frac{1}{2}$
(dashes) and spin--$\frac{3}{2}$ (dots) excited fermions both
with mass $M_{1/2(3/2)} = 250$ GeV at $\protect\sqrt{s} = 500$
GeV with unpolarized beams. We considered  coupling strengths
that lead to  deviations of $3\sigma$ in the total cross
section with respect to the standard model prediction (solid).}
\label{fig:dsig:dm}
\end{figure}

The existence of an excited fermion with mass below the
kinematical reach of the $e^-\gamma$ machine can be established
through the identification of its Breit--Wigner profile in the
$e^-\gamma$ invariant mass distribution ($M$), which should be an
easy task even in the case of unpolarized beams.  We present in
Fig.\ \ref{fig:dsig:dm} the distribution $d\sigma/dM$ for
$M_{1/2(3/2)} = 250$ GeV at an $e^+e^-$ collider with $\sqrt{s} =
500$ GeV, where we introduced a cut in  the polar angle ($\theta$)
of the final state particles with the beam pipe requiring that
$5^\circ < \theta < 175^\circ$.  For the sake of comparison, we
plotted this distribution for QED ({\it i.e.\/} $\Lambda=
\infty$) and for  values of the compositeness scale that lead to
a $3\sigma$ effect in the total cross section.

The search for excited leptons certainly can be conducted using
unpolarized beams, however, polarization can be used not only to expand
the discovery region in the $\Lambda\times M_{1/2(3/2)}$ plane, through
the enhancement of the luminosity and the cross section, but also to
study in detail the interaction of the new states. As we pointed out
before, the distribution functions assume approximately the same value
at $\bar{x} = \zeta / (\zeta +2) \simeq 0.71$, even for different
polarization configurations of the initial particles. In the interval
$0< x < \bar{x}$, the luminosity is higher for $P_p P_l > 0$, whereas
for the range $x > \bar{x}$ the distribution with $P_p P_l < 0$
dominates. Therefore, in order to search for excited leptons with mass
below (above) $\overline{M}=\sqrt{\bar{x}s}$, we should employ the
polarization configurations of the electron and the laser in such a way
that $P_p P_l > 0 \; (<0)$.  In both cases, the degree of circular
polarization of the scattered photon ($\xi_2$) has the same sign as the
positron polarization in the region of interest. Moreover, we can
see from the helicity amplitudes presented in the Appendix that
$\xi_2 >0$ enhances the cross section for  RH spin--$\frac{1}{2}$
and LH spin--$\frac{3}{2}$ excited states, while $\xi_2 < 0$
favors those with LH spin--$\frac{1}{2}$ and RH
spin--$\frac{3}{2}$.  This behavior can be easily traced to the
angular momenta configuration of the initial state.

Keeping the above comments in mind, we can envisage four different
scenarios to enlarge the discovery region in the $\Lambda\times
M_{1/2(3/2)}$ plane, depending on the spin, mass, and couplings of the
excited fermion:

($i$) $P_p >0 $ and $ P_l > 0$ for RH spin--$\frac{1}{2}$ or LH
spin--$\frac{3}{2}$ excited states with mass below $\overline{M}$;

($ii$) $P_p < 0 $ and $ P_l < 0$  for LH spin--$\frac{1}{2}$ or RH
spin--$\frac{3}{2}$ excited states with mass below $\overline{M}$;

($iii$) $P_p <0$ and  $P_l >0$ for LH spin--$\frac{1}{2}$ or RH
spin--$\frac{3}{2}$ excited states with  mass above $\overline{M}$;

($iv$) $P_p >0$ and  $P_l <0$ for RH spin--$\frac{1}{2}$ or LH
spin--$\frac{3}{2}$ excited states with  mass above $\overline{M}$.

Besides the above procedure, we can also increase the excited fermion
cross section by polarizing the electron beam. In fact,  we can
learn from the expressions for the subprocess cross section given in the
Appendix, that the use of negatively (positively) polarized electrons
enhances the signal for excited states with LH (RH) couplings, as is
naively expected. In general, polarized electron beams increase  the
sensitivity to $\Lambda$ by a factor $2$--$3$ for a given value  of
$M_{1/2(3/2)}$.

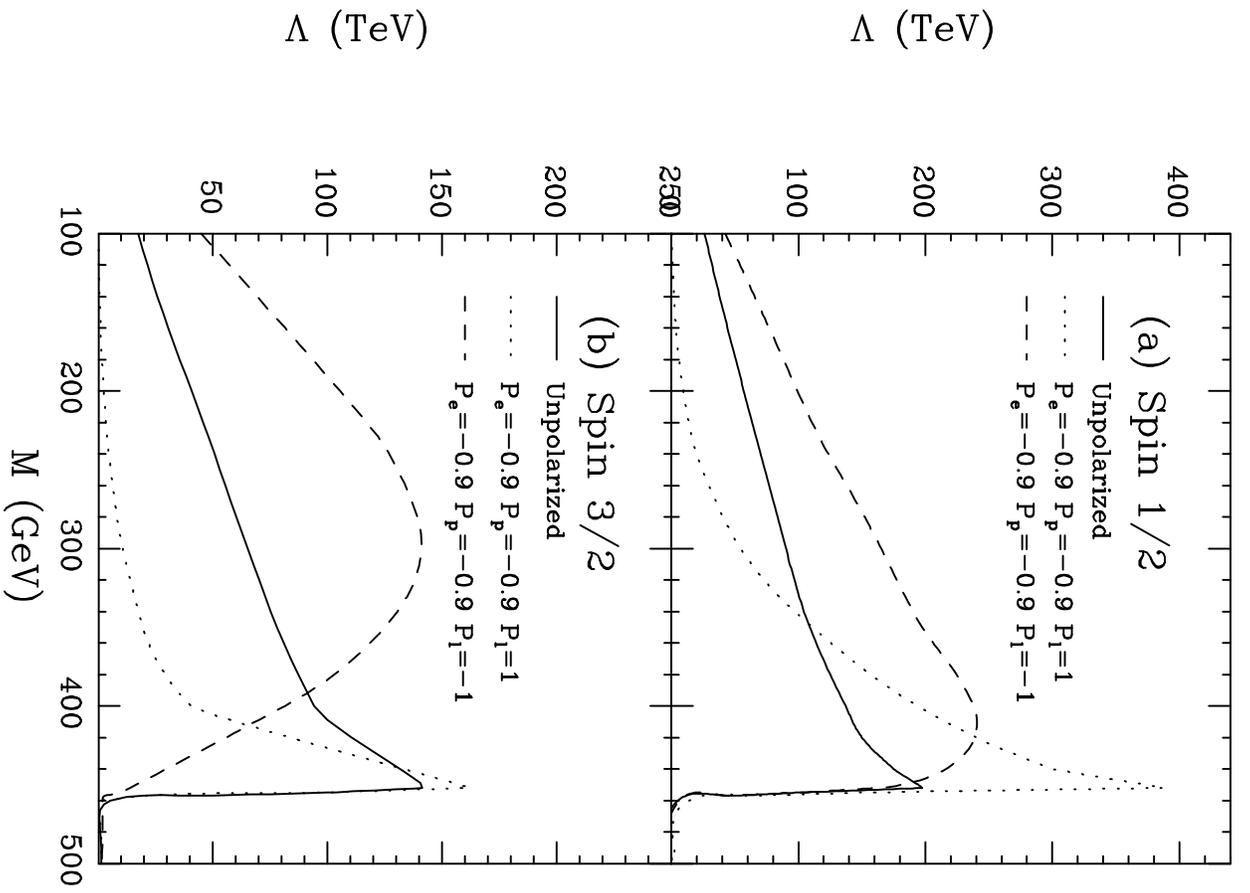
\begin{figure}
\protect
\caption{(a) Discovery contour for LH excited states of spin--$\frac 12$
for  $\protect\sqrt{s}=500$ GeV and ${\cal L}=10$ $\mbox{fb}^{-1}$. The
dotted line corresponds to $P_{e} = -0.9$,  $P_{p} = -0.9$, and $P_l =
1$ while the dashed one stands for $P_{e} = -0.9$, $P_{p} = -0.9$, and $P_l
= -1$. (b) The same as (a) for  LH spin--$\frac 32$ where the dotted
line corresponds to $P_{e} = -0.9$,  $P_{p} = 0.9$,  and $P_l = -1$ and
the dashed one to $P_{e} = -0.9$, $P_{p} = 0.9$, and $P_l = 1$. In both
figures, the solid lines stand for unpolarized beams.  By taking the
opposite of all polarizations, the same discovery contours  are valid
for RH couplings.}
\label{fig:disc:pol12}
\label{fig:disc:pol32}
\end{figure}

Figs.\ \ref{fig:disc:pol12}a and \ref{fig:disc:pol32}b show the
discovery region for several polarizations of the initial particles,
where the region that can be accessed at the NLC is located below and to
the left of the curves. For excited electron masses lower then the
kinematical limit of the $e^-\gamma$ collider, we required a $3\sigma$
effect in the cross section obtained by the integration over a bin of
$5$ GeV around the excited lepton mass. We also introduced the angular
cut $5^\circ < \theta < 175^\circ$, that mimics the angular coverage  of
a detector and also reduces the size of the QED background. For masses
larger than the kinematical limit of the collider, we evaluated ${\cal
S}$ using the total cross section. We can learn from these figures that
electron--photon collisions are able to explore very large values of the
compositeness scale, extending considerably the limits currently
available from low energy experiments. For instance, in the case of an
excited spin--$\frac{1}{2}$ lepton with mass around 400 GeV we can probe
up to $\Lambda \sim 200$ TeV.

Once the existence of excited electrons is established, it is important
to study its spin and couplings.  At this point the use of polarized
beams is crucial to determine the properties of the excited fermion. In
order to unravel the handness of the excited electron coupling,  we
studied the cross section (\ref{cross:pol}) when only the electron beam
is polarized. In this case, we can write
\begin{equation}
\sigma = \sigma_{00}
\left(1 + P_e {\cal A}_{LR} \right) \; ,
\label{sig:pe0}
\end{equation}
where $\sigma_{00}$ stands for the cross section for unpolarized beams
and ${\cal A}_{LR}$ is the left--right asymmetry factor
\begin{equation}
{\cal A}_{LR} = \frac{ \sigma_{R} - \sigma_{L} }
{ \sigma_{R} + \sigma_{L} } \; ,
\label{ass:pe}
\end{equation}
with $\sigma_{R(L)}$ being the cross section for fully polarized
electron beams and unpolarized backscattered photons.

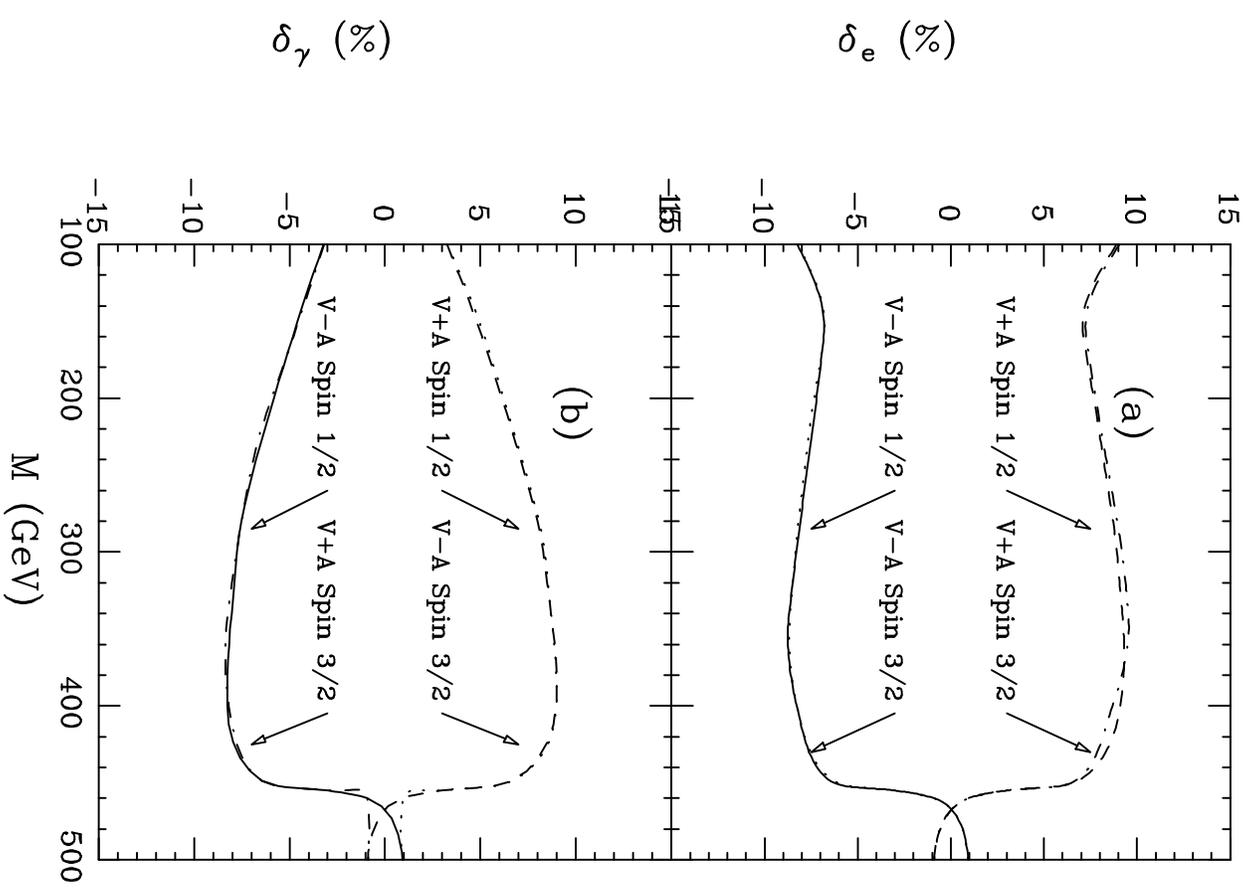
\begin{figure}
\protect
\caption{(a) Deviation $\delta_e$ [Eq.\
(\protect\ref{delta:pe})], at  $\protect\sqrt{s} = 500$ GeV, for
LH spin$\frac 12$ (solid), RH spin$\frac 12$ (dashes),
LH spin$\frac 32$ (dots), and RH spin$\frac 32$ (dotdash) excited
states as function of its mass for $P_{e} = 0.9$ and unpolarized
backscattered photons.  (b) Deviation $\delta_\gamma$ [Eq.\
(\protect\ref{delta:pg})] for unpolarized electron and laser
beams and a positron beam with $P_{p} = 0.9$.  In both figures,
these deviations have an statistical  significance of $3\sigma$.}
\label{fig:ass:pe}
\label{fig:ass:pg}
\end{figure}

In Fig.\ \ref{fig:ass:pe}a we exhibit the deviation
\begin{equation}
\delta_e = \frac{\sigma_{e0} - \sigma_{00}}{\sigma_{00}}
\label{delta:pe}
\end{equation}
of the polarized cross section ($\sigma_{e0}$) with respect to
the unpolarized one ($\sigma_{00}$), as
function of the excited state mass. Our results were obtained
integrating the cross sections over a $5$ GeV bin around the resonance
and assuming that the electron beam has a degree of longitudinal
polarization of 90\%. We required that the total number of events, for
the polarized case, differs by 3$\sigma$ from the unpolarized
yield. From this figure, we can witness that the measurement of such a
deviation is capable of distinguishing very clearly RH from LH
couplings, as it is expected from naive arguments. Moreover, the
left--right asymmetry can also be inferred, in a straightforward way,
from this measurement. Notwithstanding, we can not discriminate
between excited fermions of spin $\frac{1}{2}$ and $\frac{3}{2}$ by
the analysis of this deviation only.  This task can be accomplished by
studying the same process with polarized photon beams.

In fact, as we have pointed above, $\xi_2 >0$ enhances the cross section
for RH spin--$\frac{1}{2}$ and LH spin--$\frac{3}{2}$ excited states,
while $\xi_2 < 0$ favors LH spin--$\frac{1}{2}$ and RH
spin--$\frac{3}{2}$ excited states. Therefore, we defined the deviation
\begin{equation}
\delta_\gamma = \frac{\sigma_{0\gamma} - \sigma_{00}}{\sigma_{00}} \; ,
\label{delta:pg}
\end{equation}
where $\sigma_{00}$ is the unpolarized cross section and
$\sigma_{0\gamma}$
is the cross section integrated over the $5$ GeV bin around the
resonance for polarized backscattered photons and unpolarized electrons.
In this case, we cannot define an asymmetry factor, in the same way we did
before,  since $\xi_2$ is a function of the momentum carried by the
photon (see Eq.\ (\ref{xi:2})).

We present in Fig.\ \ref{fig:ass:pg}b the deviation $\delta_{\gamma}$
as a function of the excited state mass, requiring that the total
number of events of the polarized case differs by 3$\sigma$ from the
unpolarized one. In order to obtain the polarized backscattered
photons, we assumed that the parent positron beam has a 90\% degree of
polarization and that the laser beam is not polarized. In this setup,
the polarization of the backscattered photon ($\xi_2$) has the same
sign of the parent positron polarization for the whole spectrum.  From
this figure, we verify that RH spin--$\frac{1}{2}$ and LH
spin--$\frac{3}{2}$ fermions lead to positive values of this
deviation, whereas LH spin--$\frac{1}{2}$ and RH spin--$\frac{3}{2}$
fermions furnish negative values for it
\cite{foot2}.  Therefore, once the handness of the coupling is
established through the analysis of $\delta_{e}$ the spin of the
resonance can be determined by measuring $\delta_{\gamma}$.

\section{Electron--Positron Collisions}
\label{ee:proc}

We learned in the previous section that an $e^-\gamma$ collider is a
powerful tool to investigate the existence of excited fermions with mass
below the kinematical reach of the machine. For masses  above this
limit, it is worthwhile to employ the $e^+ e^-$ mode of the collider,
since excited fermions  appear as a virtual state being exchanged
on the $t$--channel of  the process $e^+e^- \rightarrow \, \gamma \,
\gamma$.

The polarized cross section for this process is given by
\begin{eqnarray}
d\sigma_{P_e P_p} &=&  \frac{1}{4}  \Bigl[(1+P_p P_e)
\left(d\sigma_{++} + d\sigma_{--} \right) +  (P_p +
P_e) \left(d\sigma_{++} - d\sigma_{--} \right)
\nonumber \\
&+&   (1-P_p P_e)\left(d\sigma_{+-} + d\sigma_{-+}\right)
+  (P_p - P_e)\left(d\sigma_{+-} - d\sigma_{-+} \right)  \Bigr] \; ,
\label{cross:pol:ee}
\end{eqnarray}
where $ P_{e(p)}$ is the electron (positron) mean longitudinal
polarization and $d\sigma_{\lambda _p \lambda _e}$, with  $\lambda
_{p(e)} = \pm 1$, is the cross section for fully polarized $e^+$ and
$e^-$ beams.  For completeness, the helicity amplitudes for the exchange
of a  spin--$\frac{1}{2}$ (\ref{epem12:full}) and a spin--$\frac{3}{2}$
(\ref{epem32:full}) excited electron are presented in the Appendix.

For the sake of comparison, Fig.\ \ref{c_epem} contains the angular
distribution of the produced photons in $e^+ e^- \rightarrow \gamma
\gamma$ for QED and excited leptons of spin $\frac{1}{2}$ and
$\frac{3}{2}$. The lines in this figure correspond to couplings that
lead to a 3$\sigma$ deviation from the QED total cross section.  As we
could expect, unpolarized beams cannot be used to determine the
chirality of the coupling of the excited electron. Moreover, only a
detailed experimental study of the angular distribution of the two
photons could, in principle, distinguish between a spin--$\frac{1}{2}$
and a spin--$\frac{3}{2}$ new fermion contribution.

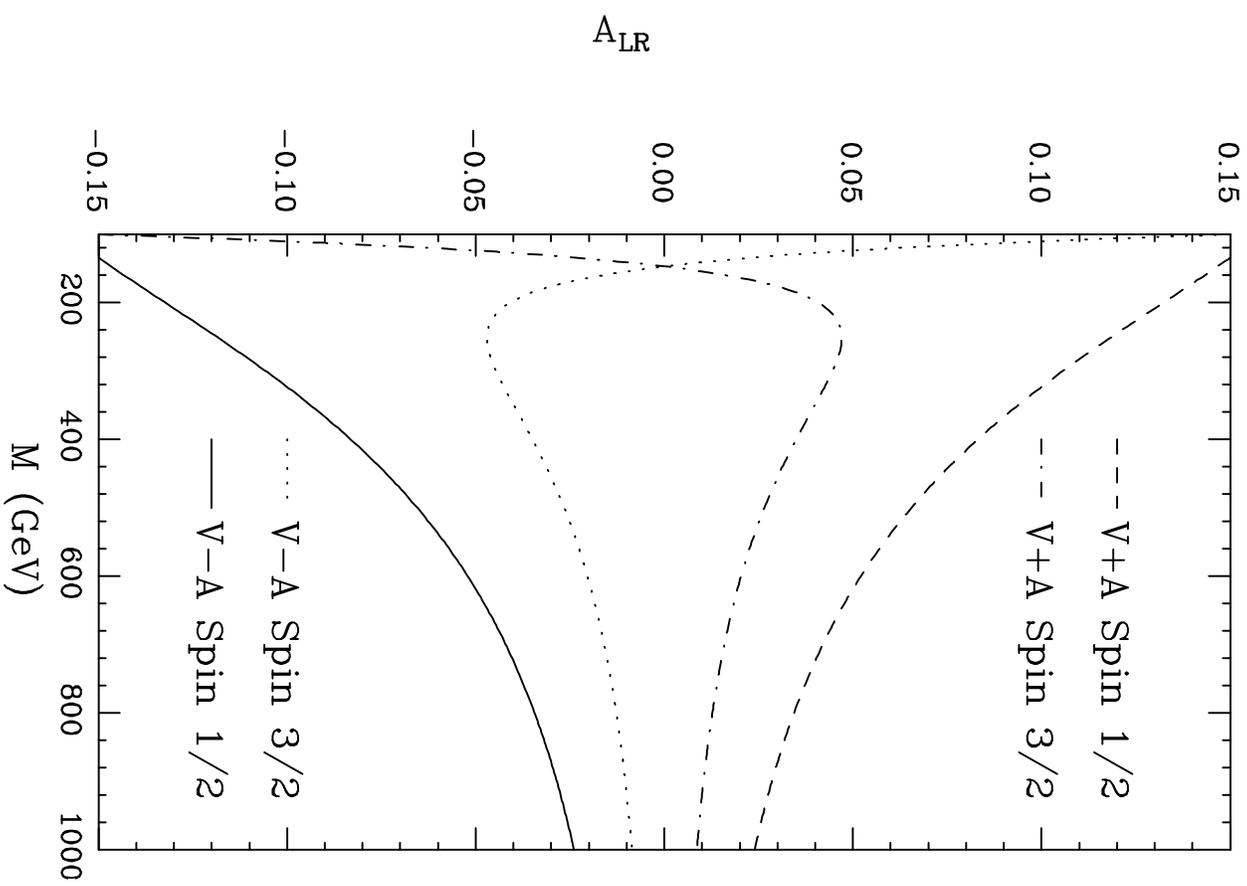
\begin{figure}
\protect
\caption{Left--right asymmetry factor ${\cal A}_{LR}$, at
$\protect\sqrt{s} = 500$ GeV, for LH spin$\frac 12$ (solid),
RH spin$\frac 12$ (dashes), LH spin$\frac 32$ (dots), and
RH spin$\frac 32$ (dotdash) excited states as function of their
masses, for $P_{e} = 0.9$ and $P_{p} = 0$.}
\label{ass:eef}
\end{figure}

Since polarization increases or reduces the total cross section
according to the chirality of the couplings, the distinction between RH
and LH couplings can be made very easily by polarizing just the electron
beam. In this case, we can compute the left--right asymmetry ${\cal
A}_{LR}$  aforementioned. In Fig.\ \ref{ass:eef}, we show ${\cal
A}_{LR}$ as function  of $M_{1/2(3/2)}$ for $\Lambda = 1$ TeV. Notice
that the chirality nature of the excited state can be clearly discerned
by looking at the sign of this asymmetry.

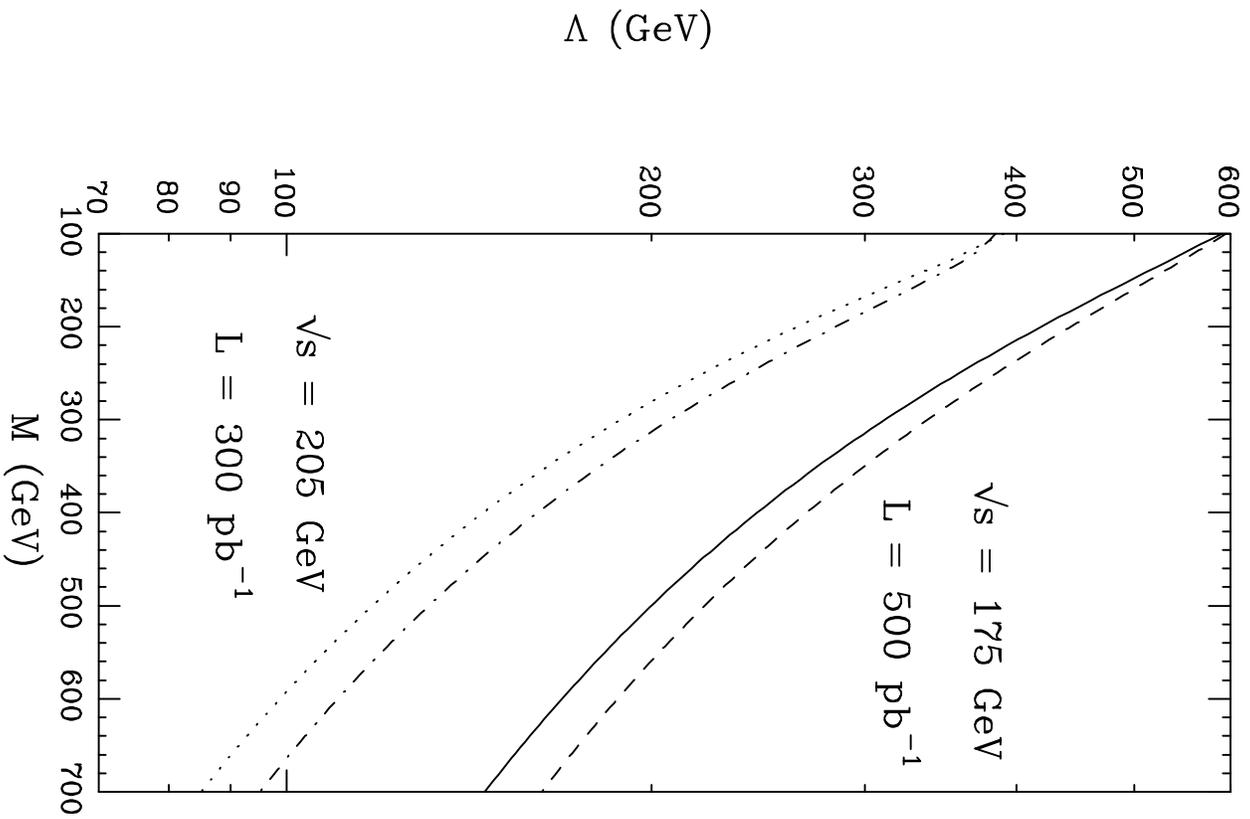
\begin{figure}
\protect
\caption{Discovery regions for LEP II. The limits for the
spin--$\frac{1}{2}(\frac{3}{2})$ excited state is represented by
the solid (dotted) line for $\protect\sqrt{s} = 175$ GeV with
${\cal L} = 500$ pb$^{-1}$, and by the dashed (dotdashed) one for
$\protect\sqrt{s} = 205$ GeV with ${\cal L} = 300$ pb$^{-1}$. We
performed an angular cut of 12$^{\circ}$ in the polar angle
of the final state photons.}
\label{disc:lep2}
\end{figure}

We present in Fig.\ \ref{disc:lep2} the discovery limits obtained from
this reaction for unpolarized beams at the LEP II energies. We
considered two different sets of parameters for this collider, {\it
i.e.\/} $\sqrt{s} = 175$ GeV with ${\cal L} = 500$ pb$^{-1}$ and
$\sqrt{s} = 205$ GeV with ${\cal L} = 300$ pb$^{-1}$. We  required a
$3\sigma$ effect in the total cross section for the $\gamma \gamma$
production and imposed a cut of $12^\circ$ in the polar angle of the
final photons with the beam pipe. We can see from this simulation that
the discovery region is larger for the $\sqrt{s} = 175$ GeV and ${\cal
L} = 500$ pb$^{-1}$ operation mode, no matter the spin of the excited
state.  As expected, both LH and RH couplings give the same result.

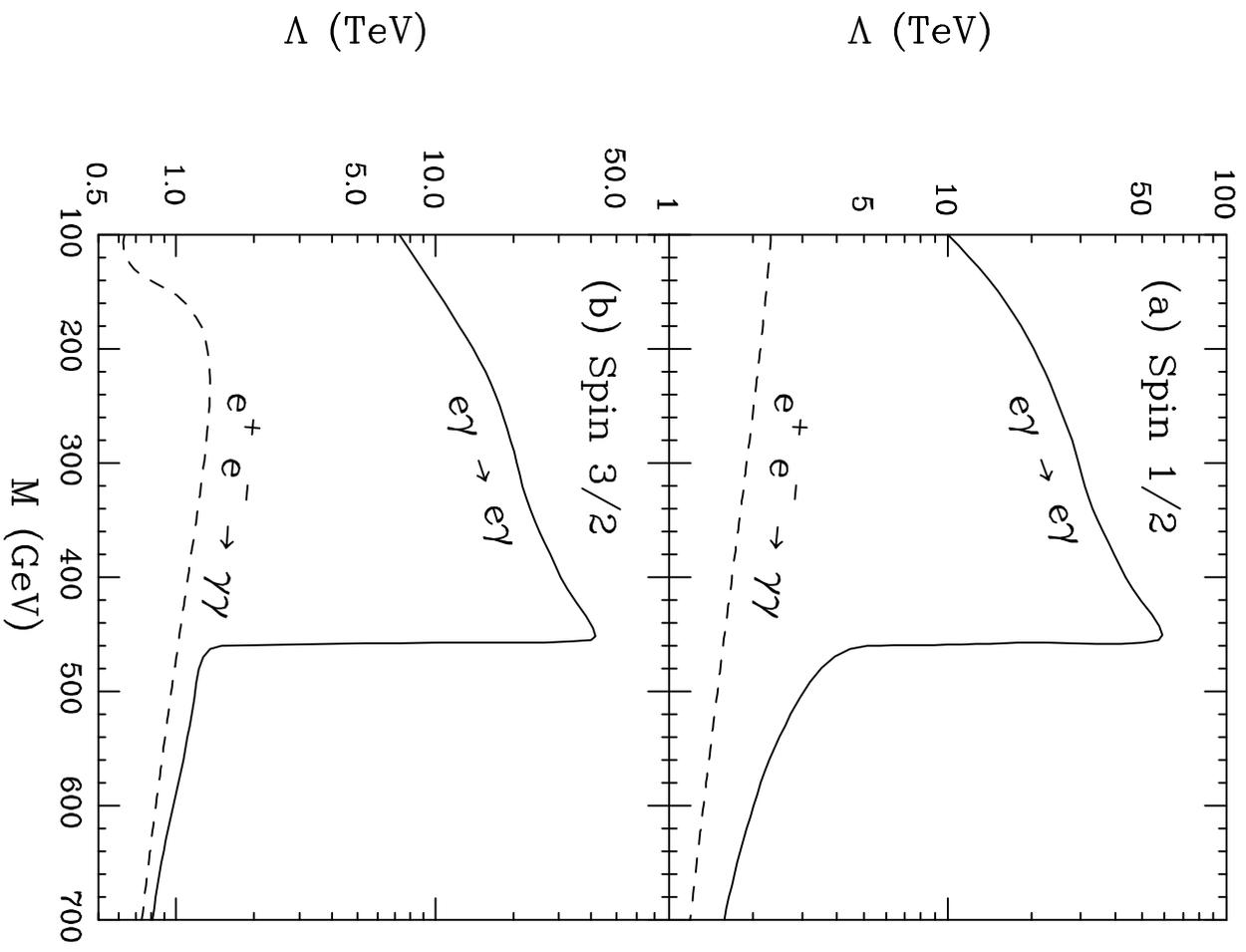
\begin{figure}
\protect
\caption{Comparison of the discovery regions for $e^+ e^-$
(dashed) and $e^- \gamma$ (solid) operating modes with
unpolarized beams, at $\protect\sqrt{s} = 500$ GeV. The figure a
(b) shows the discovery regions for spin$\frac{1}{2}$
($\frac{3}{2}$)excited states.}
\label{disc:comp}
\end{figure}

Finally, we compare in Fig.\ \ref{disc:comp}a (\ref{disc:comp}b) the
discovery regions of spin--$\frac{1}{2}$($\frac{3}{2}$) excited states
for the machine operating in the $e^-\gamma$ and $e^+ e^-$ modes.  We
considered $3\sigma$ deviations in the total cross section and assumed
unpolarized beams, performing a 5$^\circ$ cut on the polar angle to
avoid the beam pipe region. As we can see from these figures, the
$e^-\gamma$ operating mode is by far more advantageous to investigate
the existence of such states even when their masses are larger than
the center--of--mass energy of the machine.

\section{Conclusion}
\label{concl}

The existence of excited fermions is a direct consequence of a possible
new layer of matter. In this paper we have analyzed the capability of an
$e^+ e^-$ machine, operating both in $e^+ e^-$ and $e^- \gamma$ modes,
to discover and study such new states. Using polarized beams, we showed
how to determine whether the spin of the excited state is $\frac{1}{2}$
or $\frac{3}{2}$ and also the chiral structure of its coupling to
photons and usual fermions.

We showed that $e^-\gamma$ reaction is the best way to search for
excited electrons even for masses above the kinematical reach of the
machine.  Moreover, in these collisions, an important r\^ole is played
by the polarization of the beams since it allows the identification of
the chiral structure of the excited state coupling through the
measurement of deviations from the unpolarized cross section,  when
only the electron beam is polarized. The identification of its spin can
be done in the $e^-\gamma$ mode when, in addition to the measurement of
such deviation, a second measurement is made employing just a polarized
photon beam.  We have also showed in the $\Lambda\times M_{1/2(3/2)}$
plane how the use of polarized beams can enlarge in a significant way
the reach of these machines.


\acknowledgments

This work was partially supported by Conselho Nacional de
Desenvolvimento Cient\'{\i}fico e Tecnol\'ogico (CNPq) and by
Funda\c{c}\~ao de Amparo \`a Pesquisa do Estado de S\~ao Paulo
(FAPESP).


\appendix
\section*{}

In this appendix we present the helicity amplitudes for the
decay $L^-_{1/2(3/2)} \rightarrow e^- \gamma$ and for the
processes $e^- \gamma \rightarrow \left( e^- ,  L^-_{1/2(3/2)}
\right) \rightarrow e^- \gamma$ and $e^+ e^- \rightarrow
\gamma\gamma $, considering the couplings described by the
Lagrangians (\ref{l:12}) and (\ref{l:32}).  In order to evaluate
these helicity amplitudes we used the Weyl--van der Waerden
spinor technique for spin--$\frac{1}{2}$ \cite{hel:12} and
spin--$\frac{3}{2}$ \cite{hel:32} fermions. In this formalism the
usual propagator of spin--$\frac{3}{2}$ fermions
\[
P^{\mu\nu} (k) =  \frac{\not k + M_{3/2}}{k^2 - M_{3/2}^2}
\left[ - g^{\mu\nu} + \frac{1}{3}  \gamma^\mu \gamma^\nu +
\frac{1}{3 M_{3/2}^2} \left( \not k \gamma^\mu k^\nu +
\gamma^\nu k^\mu \not k \right) \right]  \; ,
\]
can be written in spinorial notation as\cite{foot3}
\begin{eqnarray}
P^{\mu\nu} (k) =
\frac{1}{\frac{1}{2} \{ K,\bar{K} \} - M_{3/2}^2 }
\left(\begin{array}{cc}
M_{3/2} \delta_p^{\;\;b} \; (A_{d\dot{e}\dot{f}g})_{b}^{\;\;c}
&  K_{a\dot{b}} \;
(B_{d\dot{e}\dot{f}g})^{\dot{b}}_{\;\; \dot{c}} \\
\bar{K}^{\dot{a}b} \; (A_{d\dot{e}\dot{f}g})_{b}^{\;\;c}
& M_{3/2}
\delta^{\dot{a}}_{\;\;\dot{b}} \;
(B_{d\dot{e}\dot{f}g})^{\dot{b}}_{\;\; \dot{c}}
\end{array} \right)
\sigma^{\mu d \dot{e}} \bar{\sigma}^{\nu \dot{f} g}
\; ,
\label{prop:32}
\end{eqnarray}
where
\begin{eqnarray}
(A_{d\dot{e}\dot{f}g})_{b}^{\;\;c} &=&  -
\frac{1}{2} \delta_{b}^{\;\;c}
\epsilon_{dg} \epsilon_{\dot{e}\dot{f}} -
\frac{1}{3}  \delta_{g}^{\;\;c}
\epsilon_{bd} \epsilon_{\dot{e}\dot{f}}
+ \frac{1}{6 M_{3/2}^2}
\left( \delta_{d}^{\;\;c}  K_{b\dot{e}} K_{g\dot{f}}
- \epsilon_{bg}  K_{d\dot{e}} K^{c}_{\;\;\dot{f}} \right) \; ,
\nonumber \\
(B_{d\dot{e}\dot{f}g})^{\dot{b}}_{\;\; \dot{c}} &=&
-  \frac{1}{2}
\delta^{\dot{b}}_{\;\;\dot{c}}  \epsilon_{dg}
\epsilon_{\dot{e}\dot{f}}
+ \frac{1}{3}  \delta^{\dot{b}}_{\;\;\dot{e}}
\epsilon_{dg}
\epsilon_{\dot{c}\dot{f}}
+ \frac{1}{6 M_{3/2}^2}
\left( \delta^{\dot{b}}_{\;\;\dot{f}}
\bar{K}_{\dot{c}g} \bar{K}_{\dot{e}d}
- \epsilon_{\dot{c}\dot{e}}
\bar{K}^{\dot{b}}_{\;\;d} \bar{K}_{\dot{f}g}
\right) \; .
\label{prop:AB}
\end{eqnarray}

In the decay $L^-_{1/2(3/2)} \rightarrow e^- \gamma$ we denote
the square of the helicity amplitudes for the decay of a
spin--$\frac{1}{2}$ excited fermion by  $|{\cal M}_{\lambda_L ;
\lambda_e \lambda_\gamma }^{1/2}|^2$, where $\lambda_L$ is the
excited lepton spin in the direction of quantization and
$\lambda_{e(\gamma)}$ the final electron (photon) helicity. It is
easy to obtain the non--zero amplitudes
\begin{eqnarray}
|{\cal M}_{+;++}^{(1/2)}|^2 & = & \frac{2e^2}{\Lambda^2} (A+B)^2
M_{1/2}^4 \; ,
\nonumber \\
|{\cal M}_{-;--}^{(1/2)}|^2 & = & \frac{2e^2}{\Lambda^2} (A-B)^2
M_{1/2}^4 \; .
\end{eqnarray}
For the decay of a spin--$\frac{3}{2}$ excited fermion, following
the notation of Novaes and Spehler \cite{hel:32}, we have
\begin{eqnarray}
|{\cal M}_{++;+-}^{(3/2)}|^2 & = & \frac{e^2}{\Lambda^2} (A+B)^2
M_{3/2}^4 \; ,
\nonumber \\
|{\cal M}_{--;-+}^{(3/2)}|^2 & = & \frac{e^2}{\Lambda^2} (A-B)^2
M_{3/2}^4 \; .
\end{eqnarray}

In the reaction $e(p^\mu)\gamma(q^\mu) \rightarrow
e(k^\mu)\gamma(l^\mu)$, we denote the square of the amplitudes
for the exchange of a spin--$\frac{1}{2}$ (spin--$\frac{3}{2}$)
excited fermion by $|{\cal M}_{\lambda_e \lambda_\gamma
\lambda^\prime_e \lambda^\prime_\gamma}^{(1/2, 3/2)}|^2$, where
$\lambda_{e(\gamma)}$ is the initial and
$\lambda^\prime_{e(\gamma)}$ the final electron and (photon)
helicity.  The Mandelstam's variables are defined as
\begin{equation}
s = (p + q)^2 \; ,  \;\;\;\;\;
t = (p - k)^2 \; , \;\;\;\;\;
u = (p - l)^2 \; .
\end{equation}

For the exchange of a spin--$\frac{1}{2}$ excited
fermion we have
\begin{eqnarray}
%
%
\left| {\cal M}_{++++}^{(1/2)} \right| ^2  &=&
-4e^4 \left[ \frac{s}{u} - \frac{2(A+B)^2}{\Lambda^2}  \frac{s^2
\left( s-M_{1/2}^2 \right) } {\left( s-M_{1/2}^2 \right) ^2
+ \left(
M_{1/2}\Gamma_{1/2} \right) ^2}  \right.
\nonumber \\
&&\left. + \frac{(A+B)^4}{\Lambda^4}
\frac{s^3\,u} {\left( s-M_{1/2}^2 \right) ^2 +\left(
M_{1/2}\Gamma_{1/2}\right) ^2} \right] \; ,
\nonumber \\
%
%
\left| {\cal M}_{++--}^{(1/2)} \right| ^2  &=&
-4e^4 \frac{t\left( A^2-B^2 \right) ^2 M_{1/2}^2}{\Lambda^4}
\nonumber \\
&& \times \frac{\left[ s\left( u-M_{1/2}^2 \right)
+ u\left( s-M_{1/2}^2 \right)
\right] ^2 + \left(u\, M_{1/2}\Gamma_{1/2} \right) ^2 }
{\left[ \left(
s-M_{1/2}^2 \right) ^2 +\left( M_{1/2}\Gamma_{1/2}\right)^2
\right]
\left( u-M_{1/2}^2 \right) ^2 }  \; ,
\nonumber \\
%
%
\left| {\cal M}_{+-+-}^{(1/2)}\right| ^2  &=&
-4e^4 \left[ \frac{u}{s} -
\frac{2(A+B)^2}{\Lambda^2} \frac{u^2}{u-M_{1/2}^2} +
\frac{(A+B)^4}{\Lambda^4}
\frac{u^3\,s}{\left( u-M_{1/2}^2\right)^2}
\right]  \; ,
\nonumber \\
%
%
\left| {\cal M}_{-+-+}^{(1/2)}\right| ^2  &=& -4e^4
\left[ \frac{u}{s} - \frac{2(A-B)^2}{\Lambda^2}
\frac{u^2}{u-M_{1/2}^2} +  \frac{(A-B)^4}{\Lambda^4}
\frac{u^3\,s}{\left( u-M_{1/2}^2\right) ^2}  \right]  \; ,
\nonumber \\
%
%
\left| {\cal M}_{--++}^{(1/2)} \right| ^2  &=&
\left| {\cal M}_{++--}^{(1/2)} \right| ^2  \; ,
\nonumber \\
%
%
\left| {\cal M}_{----}^{(1/2)} \right| ^2  &=&
-4e^4 \left[ \frac{s}{u} - \frac{2(A-B)^2}{\Lambda^2}  \frac{s^2
\left( s-M_{1/2}^2 \right) } {\left( s-M_{1/2}^2 \right)^2 +
\left( M_{1/2}\Gamma_{1/2} \right) ^2}  \right.
\nonumber \\
&& \left. + \frac{(A-B)^4}{\Lambda^4}
\frac{s^3\,u} {\left( s-M_{1/2}^2 \right) ^2 +\left(
M_{1/2}\Gamma_{1/2}\right) ^2} \right]  \; .
\label{amp:12:full}
\end{eqnarray}

In the case of spin--$\frac{3}{2}$ excited fermion we
obtain
\begin{eqnarray}
%
%
\left|  {\cal M}_{++++}^{(3/2)}\right| ^2  &=&
-4e^4 \left[ \frac{s}{u} -\frac{2\left( C+D\right) ^2}{\Lambda^2}
\frac{s^2}{6M_{3/2}^2} \frac{u+2M_{3/2}^2}{u-M_{3/2}^2}\right.
\nonumber \\
&& \left. + \frac{\left( C+D\right) ^4}{\Lambda^4}
\frac{s^3\,u}{36M_{3/2}^4} \frac{\left( u+2M_{3/2}^2\right)
^2}{\left(u - M_{3/2}^2\right) ^2 } \right]  \; ,
\nonumber \\
%
%
\left|  {\cal M}_{++--}^{(3/2)}\right| ^2  &=&
-4e^4 \frac{\left( C^2-D^2\right)^2}
{\Lambda^4}\frac{t^3}{9M_{3/2}^2}  \; ,
\nonumber \\
%
%
\left|  {\cal M}_{+-+-}^{(3/2)}\right| ^2  &=&
-4e^4 \left[ \frac{u}{s}
-  \frac{2\left( C+D\right) ^2}{\Lambda^2}  \frac{u^2}{6M_{3/2}^2}
\frac{\left( s+2M_{3/2}^2\right)\left( s-M_{3/2}^2 \right) }
{\left( s-M_{3/2}^2 \right) ^2
+\left( M_{3/2}\Gamma_{3/2}\right) ^2 } \right.
\nonumber \\
&&  \left. + \frac{\left( C+D\right) ^4}{\Lambda^4}
\frac{u^3\,s}{36M_{3/2}^4 }
\frac{\left( s+2M_{3/2}^2\right)^2}{\left( s-M_{3/2}^2 \right)^2
+\left( M_{3/2}\Gamma_{3/2}\right) ^2 } \right] \; ,
\nonumber \\
%
%
\left|  {\cal M}_{+--+}^{(3/2)}\right| ^2  &=& -4e^4
\frac{\left( C^2 - D^2 \right) ^2}{\Lambda^4}
\frac{M_{3/2}^2 t^3}{4} \frac{\left( t+2M_{3/2}^2 \right) ^2
+ \left( M_{3/2}\Gamma_{3/2} \right)^2 }
{\left[\left( s-M_{3/2}^2 \right) ^2 +
\left( M_{3/2}\Gamma_{3/2}\right) ^2 \right]
\left( u-M_{3/2}^2 \right) ^2} \; ,
\nonumber \\
%
%
\left|  {\cal M}_{-++-}^{(3/2)}\right| ^2  &=&
\left|  {\cal M}_{+--+}^{(3/2)}\right| ^2 \; ,
\nonumber \\
%
%
\left| {\cal M}_{-+-+}^{(3/2)}\right| ^2  &=& -4e^4
\left[ \frac{u}{s} -  \frac{2\left( C-D\right) ^2}{\Lambda^2}
\frac{u^2}{6M_{3/2}^2}
\frac{\left( s+2M_{3/2}^2\right)\left( s-M_{3/2}^2 \right) }
{\left( s-M_{3/2}^2 \right) ^2
+\left( M_{3/2}\Gamma_{3/2}\right) ^2 } \right.
\nonumber \\
&&  \left. + \frac{\left( C-D\right) ^4}{\Lambda^4}
\frac{u^3\,s}{36M_{3/2}^4 } \frac{\left( s+2M_{3/2}^2\right)^2}
{\left(s-M_{3/2}^2 \right)^2 +\left( M_{3/2}\Gamma_{3/2}\right)^2}
 \right] \; ,
\nonumber \\
%
%
\left| {\cal M}_{--++}^{(3/2)}\right| ^2  &=&
\left| {\cal M}_{++--}^{(3/2)}\right| ^2   \; ,
\nonumber \\
%
%
\left| {\cal M}_{----}^{(3/2)}\right| ^2  &=& -4e^4
\left[ \frac{s}{u} -\frac{2\left( C-D\right) ^2}{\Lambda^2}
\frac{s^2}{6M_{3/2}^2} \frac{u+2M_{3/2}^2}{u-M_{3/2}^2}\right.
\nonumber \\
&& \left. + \frac{\left( C-D\right) ^4}{\Lambda^4}
\frac{s^3\,u}{36M_{3/2}^4} \frac{\left( u+2M_{3/2}^2\right)
^2}{\left(u - M_{3/2}^2\right) ^2 } \right] \; .
\label{amp:32:full}
\end{eqnarray}

For the reaction $e^+(p^\mu) e^-(q^\mu) \rightarrow \gamma(k^\mu)
\gamma(l^\mu)$, we denote the
square of
the amplitudes for the exchange of a
spin--$\frac{1}{2}$ and spin--$\frac{3}{2}$ excited fermion by
$|{\cal M}_{ \lambda_{e^+} \lambda_{e^-} \lambda_\gamma
\lambda_\gamma }^{(1/2, 3/2)}|^2$,  where
$\lambda_{e^+ (e^- )(\gamma )}$ is the  positron(electron)(photon)
helicity. The Mandelstam's variables are defined as before.

For the exchange of a spin--$\frac{1}{2}$ excited
fermion, we have
\begin{eqnarray}
%
%
\left|{\cal M}_{++--}^{(1/2)}\right| ^2 & = &
4e^4 \frac{(A^2 - B^2)^2}{\Lambda ^4} M_{1/2}^2 s
\left[ \frac{t^2}{(t - M_{1/2}^2)^2} +
\frac{u^2}{(u - M_{1/2})^2} \right.
\nonumber \\
&  & \left. + \frac{2ut}{(u - M_{1/2}^2)(t - M_{1/2}^2)} \right]
\; , \nonumber \\
%
%
\left| {\cal M}_{+--+}^{(1/2)}\right| ^2 & = &
4e^4 \left[ \frac{t}{u} -  \frac{2(A - B)^2}{\Lambda ^2}
\frac{t^2}{t - M_{1/2}^2} + \frac{(A - B)^4}{\Lambda ^4}
\frac{ut^3}{(t - M_{1/2}^2)^2} \right]
\; ,\nonumber \\
%
%
 \left| {\cal M}_{+-+-}^{(1/2)}\right| ^2 & = &
4e^4 \left[ \frac{u}{t} - \frac{2(A-B)^2}{\Lambda ^2}
\frac{u^2}{u - M_{1/2}^2} + \frac{(A - B)^4}{\Lambda ^4}
\frac{tu^3}{(u - M_{1/2}^2)^2} \right]
\; , \nonumber\\
%
%
\left| {\cal M}_{-++-}^{(1/2)}\right| ^2 & = &
4e^4\left[ \frac{t}{u} - \frac{2(A+B)^2}{\Lambda ^2}
\frac{t^2}{t-M_{1/2}^2} + \frac{(A-B)^4}{\Lambda ^4}
\frac{ut^3}{(t-M_{1/2}^2)^2} \right]
\; , \nonumber\\
%
%
\left| {\cal M}_{-+-+}^{(1/2)}\right| ^2 & = &
4e^4\left[ \frac{u}{t} - \frac{2(A+B)^2}{\Lambda ^2}
\frac{u^2}{u-M_{1/2}^2}+\frac{(A-B)^4}{\Lambda ^4}
\frac{tu^3}{(u-M_{1/2}^2)^2} \right]
\; , \nonumber\\
%
%
\left| {\cal M}_{--++}^{(1/2)}\right| ^2 & = &
\left| {\cal M}_{++--}^{(1/2)}\right| ^2
\; ,
\label{epem12:full}
\end{eqnarray}
and for the spin--$\frac 32$ exchange we have
\begin{eqnarray}
%
%
\left| {\cal M}_{++++}^{(3/2)}\right| ^2 & = &
4e^4 \frac{(C^2-D^2)^2}{\Lambda ^4}
\frac{M_{3/2}^2s^3}{4} \left[ \frac{1}{(t-M_{3/2}^2)^2}
+\frac{1}{(u-M_{3/2}^2)^2} \right. \nonumber\\
&  & \left. +\frac 2{(t-M_{3/2}^2)(u-M_{3/2}^2)} \right]
\; , \nonumber\\
%
%
\left| {\cal M}_{++--}^{(3/2)}\right| ^2 & = &
4e^4 \frac{(C^2-D^2)^2}{\Lambda ^4} \frac{s^3}{9M_{3/2}^2}
\nonumber\\
%
%
\left| {\cal M}_{+-+-}^{(3/2)}\right| ^2 & = &
4e^4\left[ \frac{t}{u} - \frac{2(C-D)^2}{\Lambda ^2}
\frac{u^2(t+2M_{3/2}^2)}{6M_{3/2}^2(t-M_{3/2}^2)} \right.
\nonumber \\
&  & \left. + \frac{(C-D)^4}{\Lambda ^4}
\frac{u^3t(t+2M_{3/2}^2)^2}{36M_{3/2}^4(t-M_{3/2}^2)^2} \right]
\; , \nonumber\\
%
%
\left| {\cal M}_{+--+}^{(3/2)}\right| ^2 & = &
4e^4\left[ \frac{t}{u} - \frac{2(C-D)^2}{\Lambda ^2}
\frac{t^2(u+2M_{3/2}^2)}{6M_{3/2}^2(u-M_{3/2}^2)} \right.
\nonumber\\
&  & \left. + \frac{(C-D)^4}{\Lambda ^4}
\frac{t^3u(u+2M_{3/2}^2)^2}{36M_{3/2}^4(u-M_{3/2}^2)^2} \right]
\; , \nonumber\\
%
%
\left| {\cal M}_{-+-+}^{(3/2)}\right| ^2 & = &
4e^4\left[ \frac{u}{t} - \frac{2(C+D)^2}{\Lambda ^2}
\frac{u^2(t+2M_{3/2}^2)}{6M_{3/2}^2(t-M_{3/2}^2)}\right.
\nonumber\\
&  & \left. + \frac{(C+D)^4}{\Lambda ^4}
\frac{u^3t(t+2M_{3/2}^2)^2}{36M_{3/2}^4(t-M_{3/2}^2)^2} \right]
\; , \nonumber \\
%
%
\left| {\cal M}_{-++-}^{(3/2)}\right| ^2 & = &
4e^4\left[ \frac{t}{u} - \frac{2(C+D)^2}{\Lambda ^2}
\frac{t^2(u+2M_{3/2}^2)}{6M_{3/2}^2(u-M_{3/2}^2)} \right.
\nonumber\\
&  & \left. + \frac{(C+D)^4}{\Lambda ^4}
\frac{t^3u(u+2M_{3/2}^2)^2}{36M_{3/2}^4(u-M_{3/2}^2)^2} \right]
\; , \nonumber\\
%
%
\left| {\cal M}_{--++}^{(3/2)} \right| ^2 & = &
\left| {\cal M}_{++--}^{(3/2)} \right| ^2
\; , \nonumber\\
%
%
\left| {\cal M}_{----}^{(3/2)} \right| ^2 & = &
\left| {\cal M}_{++++}^{(3/2)} \right| ^2
\; .
\label{epem32:full}
\end{eqnarray}

It is interesting to notice that for right-handed and left-handed
couplings of the excited fermion the amplitudes
(\ref{epem12:full}) and (\ref{epem32:full}) with equal electron
and positron helicity receive no contribution from QED or from
the excited states. This was expected since we have neglected the
external fermion masses (electron and positron).



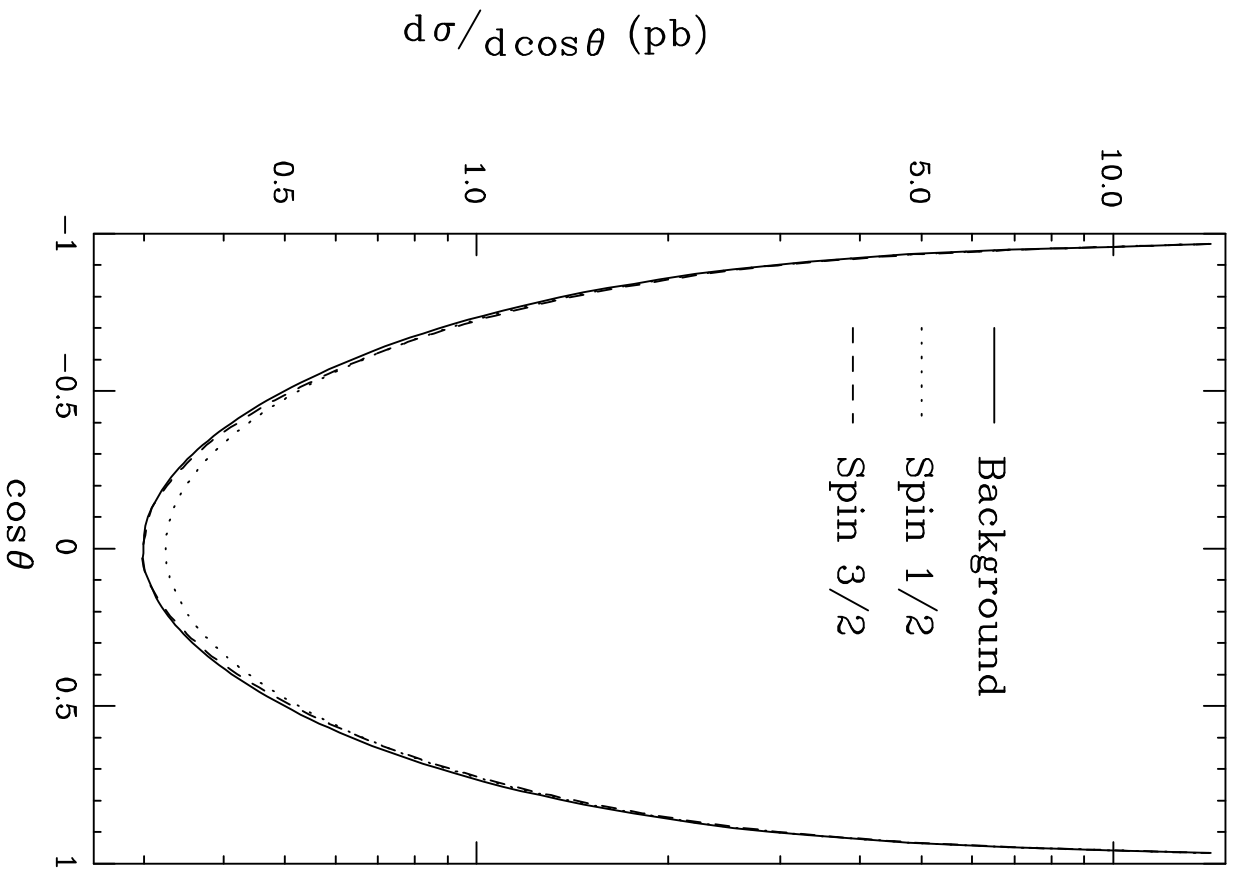
\begin{figure}
\protect
\caption{Angular distribution for the process $e^+ e^-
\rightarrow \gamma \gamma$, where $\theta$ is the angle between
the incoming electron and the outgoing photon. We assumed
unpolarized beams and  $\protect\sqrt{s} = 500$ GeV. We chose the
values of $\Lambda$ that lead to $3 \sigma$ deviations from the
standard model prediction for the total cross section. The
background is represented by the solid line and the
spin--$\frac{1}{2}$ ($\frac{3}{2}$) exchange is represented by
the dotted (dashed) line. We performed an angular cut of
5$^\circ$ on the polar angle with respect to the beam  pipe.}
\label{c_epem}
\end{figure}

\begin{references}
\bibitem{foot1} Of course, in a linear collider one can always
work with two electron beams, making it easier to obtain high
luminosity polarized beams.
\bibitem{foot2} We should notice that, for $M_{1/2(3/2)} > s$,
the interference term, that is proportional to $1/\Lambda^2$, flips
its sign, leading to a change in the sign of the deviations
$\delta_{e,\gamma}$.
\bibitem{foot3} We take this opportunity to correct some misprints
in Eq.\ (30) of Ref.\ \cite{hel:32}.
\end{references}
\end{document}